\documentclass{article}
\usepackage[utf8]{inputenc}
\usepackage{fancyhdr}
\usepackage{tikzpagenodes}
\usepackage{hyperref}
\usepackage{amsfonts}
\usepackage{amsmath}
\usepackage{cite}
\usepackage{algorithmic}
\usepackage{algorithm}
\usepackage{graphicx}
\usepackage{textcomp}
\usepackage{booktabs}
\usepackage{subcaption}
\usepackage{xcolor}
\usepackage{geometry}
\title{Data Trading with a Monopoly Social Network \\ \emph{Outcomes are Mostly Privacy Welfare Damaging}}

\date{}
\author{Ranjan~Pal, Junhui~Li,
        Yixuan~Wang,
        Mingyan~Liu,
        \thanks{R. Pal. M. Liu, J. Li, and Y. Wang are with the Department
of Electrical Engineering and Computer Science, University of Michigan, Ann Arbor, USA. E-mail: \{palr, mingyan, opheelia, joywyx\}@umich.edu,}
 Swades~De,
 \thanks{S. De is with the Department of Electrical Engineering, Indian Institute of Technology Delhi, India. E-mail: swadesd@ee.iitd.ac.in}
  and~Jon~Crowcroft 
  \thanks{J. Crowcroft is with the Computer Laboratory, University of Cambridge, UK, and the Alan Turing Institute, UK, E-mail:jac22@cam.ac.uk}}

\newtheorem{theorem}{Theorem}

\newtheorem{definition}{Definition}

\begin{document}

\maketitle
\section{abstract}
This paper argues that data of strategic individuals with heterogeneous privacy valuations in a distributed online social network (e.g., Facebook) will be under-priced, if traded in a monopoly buyer setting, and will lead to diminishing utilitarian welfare. This result, for a \textit{certain family} of online community data trading problems, is in stark contrast to a popular information economics intuition that increased amounts of end-user data signals in a data market improves its efficiency. Our proposed theory paves the way for a future (counter-intuitive) analysis of data trading oligopoly markets for online social networks (OSNs).  

\section*{keywords}
distributed community, monopoly, social welfare

\vspace{-3 mm}

\section{Introduction}
Data of billions of online individuals are currently gathered, processed, and analyzed for personalized advertising or other online service\footnote{Facebook itself has approximately 2.5 billion monthly active individual users.}. This trend is on the high rise with a perennial increase in online apps, IoT technologies, and advanced AI/ML methodologies. It is a common and age-old notion in economics (see \cite{posner1,posner2,stigler,Laudon:1996:MP:234215.234476,acquisti2016economics,odlyzko,samuelson2000privacy,schwartz2003property,posner2018radical}) that the benefits and use of sharing individual information with the demand side of an information market is beneficial to targeted customization, demand side profit, and the growth of data-`hungry' AI/ML controlled businesses. It has also been argued by economists \cite{varian2009economic,farboodi2019big} that because of the above-mentioned benefits that individual data brings to a market setting, a competitive market mechanism might generate too little data sharing from the supply side. 

In this letter, we rigorously argue, through a counter-example of a simple application type, that the popularly known economic intuition \emph{does not hold} in general, atleast for a monopoly information market setting. More specifically, we show that for some community settings (e.g., Facebook) trading end-user data/information signals in a monopoly market leads to diminishing economic utilitarian social welfare. \emph{The intuition behind this result primarily lies behind the negative externalities created via trading statistically correlated end-user signals when these heterogeneous users have varying privacy valuations of their data signals.} The result is contrary to recent results that intuit/prove that privacy can be detrimental to information market efficiency \cite{acquisti2016economics,posner2018radical,pal2019privacy, calo2015privacy,pal2020preference, acemoglu2019too, laoutaris2019online} if \emph{ideally, one's value of privacy is not high, or one's data is mildly correlated with others.} Specifically, we prove that information markets will be inefficient in \emph{non-ideal} community settings - formally hinted earlier in \cite{pal2020preference}. 
Our analysis is complete for a monopoly structure, with a major takeaway being that in practical social community settings, sub-population privacy will jeopardized at a monopoly data trading market equilibrium. To the best of our knowledge, we are the first to mathematically dispel traditional information economics intuition, albeit for social network settings only. Moreover, it paves the way for a future (counter-intuitive) analysis of general community data trading oligopoly markets. 

The rest of the paper is organized as follows. In Section II, we provide an intuition, via an example, towards proving our claim. We then follow this up in Section III with the description of a formal monopoly market model. In Section IV, we analyze this market model and formally prove our claim. We provide illustrative examples of our theory in Section V. We conclude the letter in Section VI. 

\section{Intuition}
We provide an example-driven intuition that leads us to formally investigate the validity of the hypothesis that OSN user information promotes efficient data trading markets.

We focus on the widely popular \emph{Cambridge Analytica} scandal. The company acquired private information of millions of individuals from data shared by 270,000 Facebook users who voluntarily downloaded an app for mapping personality traits, called \emph{This is your digital life}. The app accessed users' news feed, timeline, posts, and messages, and revealed information about other Facebook users. The company was finally able to infer valuable information about more than 50 million Facebook users, which it deployed for designing personalized political messages and advertising in the Brexit referendum and the 2016 US presidential election. This scandal highlighted two important facets: (i) private information (e.g., behavior, habits, preferences) of users part of an online social community such as Facebook are correlated and results in knowing such information about other users\footnote{Habits and preferences of a highly educated gay from a particular locality is informative about others with the same profile and residing in the same area.}  whose data is not leaked, and (ii) once it is openly publicized that valuable user information has been breached to satisfy external objectives, users are often miffed resulting in a huge social uproar as did happen in the case of the Cambridge Analytica scandal. These observations motivated us to develop a skepticism regarding the popular economic notion that more data implies increased information market efficiency. It could also be that trading in return for incentives for such community settings might not go down well\footnote{This is a high chance in scenarios of social uproar post publicly known data breaches, if not in cases where data breaches go un-noticed.}  with the user privacy preferences, consequently hampering societal welfare (see Section III for a definition). 

To state our intuition in a relatively more formal manner (courtesy of \cite{acemoglu2019too}), consider a community platform with two users, $i$ = 1, 2. Each user owns its own personal data, which we represent with a random variable $X_{i}$ (from the viewpoint of the platform). The relevant data of the two users are related, which we capture by assuming that their random variables are jointly normally distributed with mean zero and correlation coefficient $\rho$. The community platform can acquire or buy the data of a user in order to better estimate her preferences or actions. Its objective is to minimize the mean square error of its estimates of user types, or maximize the amount of leaked information about them. 
Suppose that the valuation (in monetary terms) of the platform for the users’ leaked information is one, while the value that the first user attaches to her privacy, again in terms of leaked information about her, is $\frac{1}{2}$ and for the second user it is $v > 0$. We also assume that the platform makes take-it-or-leave-it offers to the users to purchase their data. In the absence of any restrictions on data markets or transaction costs, the first user will always sell her data (because her valuation of privacy, $\frac{1}{2}$, is less than the value of information to the platform, 1). But given the correlation between the types of the two users, this implies that the platform will already have a fairly good estimate of the second user’s information. Suppose, for illustration, that $\rho \simeq 1$. In this case, the platform will know almost everything relevant about user 2 from user 1’s data, and this undermines the willingness of user 2 to protect her data.
In fact, since user 1 is revealing almost everything about her, she would be willing to sell her own data for a very low price (approximately 0 given $\rho \simeq 1$). But once the second user is selling her data, this also reveals the first user’s data, so the first user can only charge a very low price for her data.
Therefore in this simple example, the community platform will be able to acquire both users’ data at approximately zero price. Critically, however, this price does not reflect the users’ valuation of privacy. When $v \le 1$, the equilibrium is efficient because data are socially beneficial in this case (even if data externalities change the distribution of economic surplus between the platform and users). However, it can be arbitrarily inefficient when $v$ is sufficiently high. This is because the first user, by selling her data, is creating a negative externality on the second user.
\section{System Model}
A simple example, such as the one aforementioned, clearly provides an intuition regarding the inefficiency of information trading in community settings with heterogeneous privacy valuations. In this section. en route to generalizing the validity (or invalidity) of our intuition, we propose a monopoly information trading market model (reproduced from \cite{acemoglu2019too})\footnote{We use the same notation for consistency purposes.} consisting of $n$ platform users and a profit-maximizing community platform (e.g., Facebook). 

We consider \(n\) community users represented by the set \(\mathcal{V}=\{1, \ldots, n\} .\) Each user \(i \in \mathcal{V}\) has a type denoted by
\(x_{i}\) which is a realization of a random variable \(X_{i} .\) We assume that the vector of random variables
\(\mathbf{X}=\left(X_{1}, \ldots, X_{n}\right)\) has a joint normal distribution \(\mathcal{N}(0, \Sigma),\)
where $\Sigma \in \mathbf{R}^{n \times n}$ is the publicly known covariance
matrix of \(\mathbf{X} .\) Let \(\Sigma_{i j}\) designate the \((i, j)\) -th entry of \(\Sigma\) and \(\Sigma_{i i}=\sigma_{i}^{2}>0\) denote the variance of
individual \(i^{\prime}\)s type. Each user has some personal data, \(S_{i}\), which is informative about its type, i.e., the `DNA' that drives the user's tastes (for example, based
on her past behavior, preferences, or contacts). We suppose that \(S_{i}=X_{i}+Z_{i}\) where \(Z_{i}\) is an
independent random variable with standard normal distribution\footnote{This has taken various forms in the information privacy literature \cite{sarwate2013signal}.}, i.e., \(Z_{i} \sim \mathcal{N}(0,1)\). For any user joining the community platform, the platform can derive additional revenue (e.g., due to benefits of targeted advertising) if it can predict the user's type. We simply assume that the community platform’s revenue from each user is a deceasing function of the mean square error of its forecast of the user’s type, minus what the platform pays to users to acquire their information. More specifically, the objective of the platform is to minimize
\begin{equation}
\sum_{i \in \mathcal{V}}
\left(\mathbb{E}
\left[\left(\hat{x}_{i}(\mathbf{S})-X_{i}\right)^{2}\right]-\sigma_{i}^{2}+p_{i}\right)
\end{equation}
where \(\mathbf{S}\) is the vector of data the platform acquires, \(\hat{x}_{i}(\mathbf{S})\) is the platform's estimate of the user's type given this information, \(-\sigma_{i}^{2}\) is included as a convenient normalization, and \(p_{i}\) denotes payments (be it explicit or implicit) to user \(i\) from the platform for their data (we ignore for simplicity any other transaction costs incurred by the platform). 

Users value their privacy, which we also model in a reduced-form manner (reflecting both pecuniary and non-pecuniary\footnote{As example, the fact that a user may receive a greater consumer surplus when the platform knows less about her or she may have a genuine demand for keeping her preferences, behavior, and information private. There may also be political and social reasons for privacy, for example, for concealing dissident activities or behaviors disapproved by some groups.} motives) as a function of the same mean square error. 
We assume, specifically, that user \(i^{\prime}\) s value of privacy is \(v_{i} \geq 0,\) and her payoff is $
v_{i}\left(
\mathbb{E}
\left[\left(\hat{x}_{i}(
\mathbf{S}
)-X_{i}\right)^{2}\right]-\sigma_{i}^{2}\right)+p_{i}
$
This expression and its comparison with objective (1) clarifies that the platform and users have potentially-
opposing preferences over information about user type. We have again subtracted \(\sigma_{i}^{2}\) as a normalization, which ensures that if the platform acquires no additional information about the user and makes no payment to it, the payoff is zero.
Critically, users with \(v_{i}<1\) value their privacy less than the valuation that the platform attaches to information about them, and thus reducing the mean square error of the estimates of their types is socially beneficial. In contrast, users with \(v_{i}>1\) value their privacy more, and reducing their mean square error is socially costly. In settings without data externalities (where data about one user have no relevance to the information about other users - an example being collection agencies not gathering addresses locations), the first group of
users should allow the platform to acquire (buy) their data, while the second group should not. A simple market mechanism based on prices for data can implement this efficient outcome, in accordance to the traditional economic notion that more information implies better market efficiency. However, the situation could be very different in the presence of data externalities (e.g., online community settings such as Facebook).

A key notion for our analysis is breached information, which captures the reduction in the mean square error of the platform’s estimate of the type of a user. When the platform has no information about user $i$, its estimate satisfies $
\mathbb{E}\left[\left(\hat{x}_{i}-X_{i}\right)^{2}\right]=\sigma_{i}^{2}$. As the platform receives data from this and other
users, its estimate improves and the mean square error declines. The notion of breached information captures this reduction in mean square error (MSE). Specifically, let \(a_{i} \in\{0,1\}\) denote the data sharing action of user \(i \in \mathcal{V}\) with \(a_{i}=1\) corresponding to sharing. Denote the the profile of sharing decisions by \(\mathbf{a}=\left(a_{1}, \ldots, a_{n}\right)\) and the decisions of agents other than \(i\) by \(\mathbf{a}_{-i} .\) We also use the notation \(\mathbf{S}_{\mathbf{a}}\) to denote the data of all individuals for whom \(a_{j}=1,\) i.e., \(\mathbf{S}_{\mathbf{a}}=\left(S_{j}: j \in \mathcal{V}\, \mathrm{s.t.}\, a_{j}=1\right) .\) Given a profile of actions \textbf{a}, the breached information of (or about) user \(i \in \mathcal{V}\) is the reduction in the MSE of the best estimator of the type
of user \(i\) :
$\mathcal{I}_{i}(\mathbf{a})=\sigma_{i}^{2}-\min _{\hat{x}_{i}(\cdot)} 
\mathbb{E}
\left[\left(X_{i}-\hat{x}_{i}\left(\mathbf{S}_{\mathbf{a}}\right)\right)^{2}\right].$

Notably, because of data externalities, breached information about user $i$ depends not just on her decisions but also on the sharing actions taken by all users. With this notion at hand, we can write the payoff of user $i$ given the price vector $p = (p_1, . . . , p_n)$
as
{\setlength\abovedisplayskip{0pt}
\setlength\belowdisplayskip{2pt}$$
u_{i}\left(a_{i}, \mathbf{a}_{-i}, \mathbf{p}\right)=\left\{\begin{array}{ll}p_{i}-v_{i} \mathcal{I}_{i}\left(a_{i}=1, \mathbf{a}_{-i}\right), & a_{i}=1 \\ -v_{i} \mathcal{I}_{i}\left(a_{i}=0, \mathbf{a}_{-i}\right), & a_{i}=0\end{array}\right.
$$}
where recall that $v_{i} \ge 0$ is user’s value of privacy. We also express the monopoly platform’s payoff more compactly as
{\setlength\abovedisplayskip{1 pt}
\setlength\belowdisplayskip{1 pt}\begin{equation}
U(\mathbf{a}, \mathbf{p})=\sum_{i \in \mathcal{V}} \mathcal{I}_{i}(\mathbf{a})-\sum_{i \in \mathcal{V}: a_{i}=1} p_{i}
\end{equation}
}
An action profile $a = (a_1,...,a_n)$ of the strategic users and a price vector $p = (p_1,...,p_n)$ for the users constitute a pure strategy equilibrium of the user-platform game if both users and the community platform maximize their payoffs given other players’ strategies. More formally, we define an equilibrium of this game as a \emph{Stackelberg equilibrium} in which the monopoly platform chooses the price vector recognizing the user equilibrium that will result following this choice.

\begin{definition}
\emph{Given the price vector \(\mathbf{p}=\left(p_{1}, \ldots, p_{n}\right),\) an action profile \(\mathbf{a}=\left(a_{1}, \ldots, a_{n}\right)\) is user equilibrium if for all \(i \in \mathcal{V}\),
{\setlength\abovedisplayskip{-2pt}
\setlength\belowdisplayskip{2pt}$$
a_{i} \in \operatorname{argmax}_{a \in\{0,1\}} u_{i}\left(a_{i}=a, \mathbf{a}_{-i}, \mathbf{p}\right).
$$}}
\end{definition}
We denote the set of user equilibria at a given price vector \(\mathbf{p}\) by \(\mathcal{A}(\mathbf{p})\). A pair \(\left(\mathbf{p}^{\mathrm{E}}, \mathbf{a}^{\mathrm{E}}\right)\) of price and
action vectors is a pure strategy Stackelberg equilibrium if \(\mathbf{a}^{\mathrm{E}} \in \mathcal{A}\left(\mathbf{p}^{\mathrm{E}}\right)\) and there is no profitable
deviation for the platform, i.e.,
{\setlength\abovedisplayskip{1pt}
\setlength\belowdisplayskip{0pt}$$
U\left(\mathbf{a}^{\mathrm{E}}, \mathbf{p}^{\mathrm{E}}\right) \geq U(\mathbf{a}, \mathbf{p}), \; {\rm for \; all \;} \mathbf{p} {\rm \;and \; for \; all \; } \mathbf{a} \in \mathcal{A}(\mathbf{p})
$$}
In what follows, we refer to a pure strategy Stackelberg equilibrium simply as an equilibrium.

We now characterize two important properties of the breached information function \(\mathcal{I}_{i}\) :
\(\{0,1\}^{n} \rightarrow \mathbf{R}\). \\
1. \emph{Monotonicity:} for two action profiles $\textbf{a}$ and $\textbf{a}'$ with $\textbf{a}$ \(\geq\) $\textbf{a}'$
{\setlength\abovedisplayskip{2pt}
\setlength\belowdisplayskip{2pt}$$
\mathcal{I}_{i}(\mathbf{a}) \geq \mathcal{I}_{i}\left(\mathbf{a}^{\prime}\right), \quad \forall i \in\{1, \ldots, n\}
$$}
2. \emph{Submodularity:} for two profiles $\textbf{a}$ and $\textbf{a}'$ with $\mathbf{a'}_{-i} \ge \mathbf{a}_{-i}$,
{\setlength\abovedisplayskip{2pt}
\setlength\belowdisplayskip{2pt}{\small
$$
\mathcal{I}_{i}\left(a_{i}=1, \mathbf{a}_{-i}\right)-\mathcal{I}_{i}\left(a_{i}=0, \mathbf{a}_{-i}\right) \geq \mathcal{I}_{i}\left(a_{i}=1, \mathbf{a}_{-i}^{\prime}\right)-\mathcal{I}_{i}\left(a_{i}=0, \mathbf{a}_{-i}^{\prime}\right)
$$
}}
The monotonicity property states that as the set of community users who share their information expands, the breached information about each user (weakly) increases. This is an intuitive consequence of the fact that more information always facilitates the estimation problem of the platform and reduces the mean square error of its estimates. More important for the rest of our analysis is the submodularity property, which implies that the marginal increase in the breached information from individual $i$’s sharing decision is decreasing in the information shared by others. This too is intuitive and follows from the fact that when others’ actions reveal more information, there is less to be revealed by the sharing decision of any given individual. Thus, from the celebrated result due to Topkis \cite{topkis1978minimizing}, for any $\mathbf{p}$, the set \(\mathcal{A}(\mathbf{p})\) is a complete lattice, and thus has a least and a greatest element. This implies that the set of user equilibria is always non-empty. 
\section{Monopoly Market Analysis}
Enroute to analyzing the market welfare generated via the aforementioned game setting, we first define the benchmark \emph{first best} welfare outcome as the data sharing decisions that maximize utilitarian social welfare or social surplus given by the sum of the payoffs of the platform and users. Social surplus (SoS) from an action profile \textbf{a} is
\[
SoS(\textbf{a})=U(\mathbf{a}, \mathbf{p})+\sum_{i \in \mathcal{V}} u_{i}(\mathbf{a}, \mathbf{p})=\sum_{i \in \mathcal{V}}\left(1-v_{i}\right) \mathcal{I}_{i}(\mathbf{a})\]

Note that prices do not appear in this expression because they are transfers from the community platform to users. The first-best action profile, $\textbf{a}^{W}$, maximizes this expression. The following theorem (built on \cite{acemoglu2019too}) characterizes the first-best action profile.

\begin{theorem}\label{Proposition-one} (due to \cite{acemoglu2019too})
\emph{The first best involves \(a_{i}^{\mathrm{W}}=1\) if
{\setlength\abovedisplayskip{1 pt}
\setlength\belowdisplayskip{1 pt}
\begin{equation}
\sum_{j \in \mathcal{V}}\left(1-v_{j}\right) \frac{\left({\rm Cov}\left(X_{i}, X_{j} | a_{i}=0, \mathbf{a}_{-i}^{\mathrm{W}}\right)\right)^{2}}{1+\sigma_{j}^{2}-\mathcal{I}_{j}\left(a_{i}=0, \mathbf{a}_{-i}^{\mathrm{W}}\right)} \geq 0
\end{equation}
}
and \(a_{i}^{\mathrm{W}}=0\) if (3) is negative.}
\end{theorem}


\textbf{Implication} - To understand this result, consider first the case in which there are no data externalities so that the covariance terms in (3) are zero, except 
 \({\rm Cov}\left(X_{i}, X_{i} | a_{i}=0, \mathbf{a}_{-i}^{\mathrm{W}}\right)=\sigma_{i}^{2}\), so that the left-hand side is simply \(\sigma_{i}^{4} /\left(1+\sigma_{i}^{2}\right)\) times $1- v_i$. This yields $a^W_i = 1$ if $v_i \le 1$ (thus a no externality setting becomes mathematically equivalent to case when all users do not value their privacy enough) . The situation is different in the presence of data externalities, because now the covariance terms are non-zero. In this case, an individual should optimally share her data only if it does not reveal too much about users with $v_j > 1$. Note here that the covariance matrix can be robustly estimated from publicly observed $S_{i}$ values as dependencies are usually preserved in the addition of noise within a threshold.  
 
In this section, we adopt the more realistic assumption that, to start with, the monopoly platform does not know the exact valuations of users in a community (in contrast to assumptions made in existing works such as \cite{wang2016value}) that are only private to them, but are informed that the valuations $v_{i}$ come from the cumulative distribution function $F_{i}$ and density function $f_{i}$ (with upper support denoted by $v_{\max}$). We allow the platform to design a pricing mechanism that elicits the true privacy valuations from the users \emph{(as Step 1 of the economy)}, somewhat similar to the seminal Vickrey-Clarkes-Groves (VCG) mechanism with a minor variation. More specifically, for any user $i \in \mathcal{V}$ the price offered to user $i$ \emph{(as Step 2 of the economy)} is equal to the surplus of all other users on the platform when user $i$ is present minus by the surplus when user $i$ is absent. We consequently have the following result.

\begin{theorem}\label{eight} (due to \cite{acemoglu2019too})
\emph{Let $\mathbf{v}$ be the reported vector of values of privacy. Then the non-negative pricing scheme}
\emph{
\begin{equation*}
p_{i}(\mathbf{v})=\left(\mathcal{I}_{i}(\mathbf{a}(\mathbf{v}))+\sum_{j \neq i}\left(1-v_{j}\right) \mathcal{I}_{j}(\mathbf{a}(\mathbf{v}))\right)
-\min _{\mathbf{a} \in\{0,1\}^{n}}\left(\mathcal{I}_{i}(\mathbf{a})+\sum_{j \neq i}\left(1-v_{j}\right) \mathcal{I}_{j}(\mathbf{a})\right)
\end{equation*}}
\emph{where $\mathbf{a}(\mathbf{v})=\arg \max _{\mathbf{a} \in\{0,1\}^{n}} \sum_{i \in \mathcal{V}}\left(1-v_{i}\right) \mathcal{I}_{i}(\mathbf{a})$ incentivizes users to report their value of privacy truthfully.}
\end{theorem}
\begin{definition}
\emph{An equilibrium is a pair \(\left(\mathbf{a}^{\mathrm{E}}, \mathbf{p}^{\mathrm{E}}\right)\) of functions of the reported valuations \(\mathbf{v}=\)
\(\left(v_{1}, \ldots, v_{n}\right)\) such that each user reports its true value and the expected payoff of the platform
is maximized. That is},
{{
\begin{equation*}
\begin{split}&\left(\mathbf{a}^{\mathrm{E}}, \mathbf{p}^{\mathrm{E}}\right)= \max _{\mathbf{a}:
\mathbb{R}^{n} \rightarrow\{0,1\}^{n}, \mathbf{p}: \mathbb{R}^{n}
\rightarrow \mathbb{R}^{n}} \mathbb{E}_{\mathbf{v}}\left[\sum_{i=1}^{n} \mathcal{I}_{i}(\mathbf{a}(\mathbf{v}))-\sum_{i: a_{i}(\mathbf{v})=1} p_{i}(\mathbf{v})\right] \\ & p_{i}(\mathbf{v})-v_{i} \mathcal{I}_{i}(\mathbf{a}(\mathbf{v})) \geq p_{i}\left(\mathbf{v}_{-i}, v_{i}^{\prime}\right)-v_{i} \mathcal{I}_{i}\left(\mathbf{a}\left(\mathbf{v}_{-i}, v_{i}^{\prime}\right)\right), \forall v_{i}^{\prime}, \mathbf{v}: i \in \mathcal{V} \end{split}
\end{equation*}}}
\end{definition}

We now have the following theorem characterizing the equilibrium of the monopoly market setting. 

\begin{theorem}\label{eleven} (due to \cite{acemoglu2019too})
\emph{For any reported vector of values v, the market equilibrium is given by}
\emph{
\begin{equation*}
\mathbf{a}^{\mathrm{E}}(\mathbf{v})=\operatorname{argmax}_{\mathbf{a} \in\{0,1\}^{n}} \sum_{i=1}^{n}\left(1-\Phi_{i}\left(v_{i}\right)\right) \mathcal{I}_{i}(\mathbf{a})
+\Phi_{i}\left(v_{i}\right) \mathcal{I}_{i}\left(\mathbf{a}_{-i}, a_{i}=0\right)
\end{equation*}
}
\emph{and}
{
\begin{equation*}
p_{i}^{\mathrm{E}}\left(v_{i}\right)= \int_{v}^{v_{\max }}
\left(\mathcal{I}_{i}\left(\mathbf{a}^{\mathrm{E}}\left(x,
\mathbf{v}_{-i})\right)-\mathcal{I}_{i}\left(\mathbf{a}_{-i}^{\mathrm{E}}\left(x, 
\mathbf{v}_{-i}
\right), a_{i}=0\right)\right)\right) d x  
+v_{i}\left(\mathcal{I}_{i}\left(\mathbf{a}^{\mathrm{E}}\left(v_{i}, \mathbf{v}_{-i}\right)\right)-\mathcal{I}_{i}\left(\mathbf{a}_{-i}^{\mathrm{E}}\left(v_{i}, \mathbf{v}_{-i}\right), a_{i}=0\right)\right)
\end{equation*}
}

\emph{Moreover, all users report truthfully and thus the expected payoff of the platform is}
{$$
\mathbb{E}_{\mathbf{v}}\left[\max _{\mathbf{a} \in\{0,1\}^{n}} \sum_{i=1}^{n}\left(1-\Phi_{i}\left(v_{i}\right)\right) \mathcal{I}_{i}(\mathbf{a})+\Phi_{i}\left(v_{i}\right) \mathcal{I}_{i}\left(\mathbf{a}_{-i}, a_{i}=0\right)\right]
$$}
\emph{where, $\Phi_{i}(v) = v + \frac{F_{i}(v)}{f_{i}(v)}$ is a non-decreasing function representing the additional rent that a user will capture in incentive compatible mechanisms.}
\end{theorem}

\textbf{Implication} - The theorem guarantees the existence of a unique monopoly market equilibrium where community platform users elicit their true valuations. A sufficient condition for $\Phi_{i}(v)$ to be non-decreasing is for  $\frac{f_{i}(v)}{F_{i}(v)}$ to be non-increasing. This requirement is satisfied for a variety of distributions such as uniform and exponential \cite{burkschat2014reversed}. 

We now investigate whether the reachable market equilibrium is efficient. We have the following result, via \cite{acemoglu2019too}, in this regard.

\begin{theorem}\label{twelve} (due to \cite{acemoglu2019too})
\emph{1. Suppose high-value users are uncorrelated with all other users and $\mathcal{V}^{(l)}=\mathcal{V}_{\Phi}^{(l)}$, where $\mathcal{V}_{\Phi}^{(l)}=\left\{i \in \mathcal{V}: \Phi_{i}\left(v_{i}\right) \leq 1\right\}$ denotes the set of users with $\Phi_{i}(v_{i}) \le 1$. Then
the market equilibrium is \textbf{efficient}.\\
2. Suppose some high-value users (those in $\mathcal{V}^{(h)}$) are correlated with users in $v_{\phi}^{(l)}$. Then there exists $\overline{\mathfrak{v}} \in \mathbb{R}^{| V(n) |}$ such that for $\mathbf{v}^{(h)} \geq \overline{\mathbf{v}}$ the market equilibrium is \textbf{inefficient}.\\
3. Suppose every high-value user is uncorrelated with all users in $\mathcal{V}_{\phi}^{(l)},$ but users in a nonempty subset
$\hat{\mathcal{V}}^{(l)}$ of $\mathcal{V}^{(l)} \backslash \mathcal{V}_{\phi}^{(l)}$ are correlated with at least one high-value user. Then there exist $\overline{\mathbf{v}}$ and $\tilde{v}$ such that if $\mathbf{v}^{(h)} \geq \overline{\mathbf{v}}$ and $v_{i}<\tilde{v}$ for some $i \in \hat{\mathcal{V}}^{(l)}$, the market equilibrium is \textbf{inefficient}.\\
4. Suppose every high-value user is uncorrelated with all low-value users and at least one high-value user is correlated with another high-value user. Let $\tilde{\mathcal{V}}^{(h)} \subseteq \mathcal{V}^{(h)}$ be the subset of high-value users correlated with at least one other high-value user. Then for each $i \in \tilde{\mathcal{V}}^{(h)}$ there exists $\bar{v}_{i}>0$ such that if for any $i \in \tilde{\mathcal{V}}^{(h)} v_{i}<\bar{v}_{i},$ the market equilibrium is \textbf{inefficient}.\\
5.The social surplus at market equilibrium, $\mathbf{a}^{\mathrm{E}}$, for any \textbf{v} (either known truthfully or otherwise)
\[SoS\left(\mathbf{a}^{\mathrm{E}}\right) \leq \sum_{i: v_{i} \in \mathcal{V}^{(l)}}\left(1-v_{i}\right) \mathcal{I}_{i}(\mathcal{V})-\sum_{i: v_{i} \in \mathcal{V}^{(h)}}\left(v_{i}-1\right) \mathcal{I}_{i}\left(\mathcal{V}_{\phi}^{(l)}\right)\]}
\end{theorem}

\textbf{Implication} - The theorem provides the conditions under which market equilibrium in a monopoly community information trading setting is utilitarian welfare (in)efficient. Note that the market efficiency results in the theorem are conservative in the sense we assume user privacy valuations are unknown in the worst case, and the platform does its best to elicit true valuation responses. Inefficiency in this setting would imply inefficiency in the case where community user valuations are untruthful (see point \#5 in the theorem). According to the points in the theorem, it is evident that the information trading market is efficient is when high-value users (those with both $v_{i}, \Phi_{i}(v_{i}) \ge 1$ are uncorrelated with all other users - something practically rare to achieve, and low-value users have $\Phi_{i}(v_{i})$ values less than one (note here that low-value users always have $v_{i} <1$ but could have $\Phi_{i}(v_{i})$ values greater than 1). Not satisfying either will lead to incentive compatibility conditions preventing efficient allocation. In all other cases, the information trading market is inefficient (SoS at equilibrium is not optimal), and the extent of inefficiency depends on whether high-value users are correlated with low-value users with $\Phi_{i}(v_{i})$ values greater or less than one.

\section{Examples}
In this section, we provide numerical examples (as in \cite{acemoglu2019too}) to lucidly illustrate (a) the existence of a data trading market equilibrium, and (b) the social surplus (SoS) zone at market equilibrium. 

\noindent\textbf{Example 1. \label{example1}}
Suppose there are two users 1 and 2 with covariance matrix
$\Sigma=\left(\begin{array}{lr}1 & \rho \\ \rho & 1\end{array}\right)$
and \(v_{1}=v_{2}=v .\) 
When \(p_{1}, p_{2} \in\left[\frac{\left(2-\rho^{2}\right)^{2}}{2\left(4-\rho^{2}\right)}, \frac{1}{2}\right],\) both action profiles \(a_{1}=a_{2}=0\) and \(a_{1}=a_{2}=1\) are user
equilibria. This is a consequence of the submodularity of the leaked information function
: when user 1 shares her data, she is also revealing a lot about user 2 , and making it less costly
for her to share her data. Conversely, when user 1 does not share, this encourages user 2 not to
share. Despite this multiplicity of user equilibria, there exists a unique (Stackelberg) equilibrium
for this game given by \(a_{1}^{\mathrm{E}}=a_{2}^{\mathrm{E}}=1\) and \(p_{1}^{\mathrm{E}}=p_{2}^{\mathrm{E}}=\frac{\left(2-\rho^{2}\right)^{2}}{2\left(4-\rho^{2}\right)} .\) This uniqueness follows because the platform can choose the price vector to encourage both users to share. \emph{The next example suggests that though there may be multiple equilibria in the Stackelberg game, all of them yield the same payoff for the community platform.} 

\noindent\textbf{Example 2. \label{example2}}
Suppose there are three users with the same value of privacy and variance: \(v_i = 1.18\) and \(\sigma_{i}^{2}=1\) for \(i=1,2,3 .\) We let all off-diagonal entries of \(\Sigma\) to be \(0.3 .\) Any action profile where two out of three users share their information is an equilibrium, and thus there are three distinct equilibria. But it is straightforward to verify that they all yield the same payoff to the platform.

The following example illustrates the social welfare zone at market equilibrium with variations in the correlation coefficient $\rho$ and privacy valuation for high-value community users. 

\noindent\textbf{Example 3. \label{example4}}
We consider a setting with two communities, each of size 10. Suppose that all users in
community 1 are low-value and have a value of privacy equal to 0.9, while all users in community
2 are high-value (with \(v_{h}>1\) ). We also take the variances of all user data to be 1, the correlation
between any two users who belong to the same community to be 1/20, and the correlation be-
tween any two users who belong to different communities to be \(\rho\).   depicts equilibrium
surplus as a function of \(v_{h}\) and \(\rho\). 

\noindent \emph{Two points are worth noting from this example. First, relatively small values of the correlation coefficient \(\rho\) are sufficient for social surplus to be negative. Second, when \(v_{h}\) is very close to 1, the social surplus is always positive because the negative surplus from high-value users is compensated by the social benefits their data sharing creates for low-value users.}
\section{Conclusion}
In this paper, we mathematically argued, using recent developments in \cite{acemoglu2019too}, that social community data trading is not economically welfare efficient in a monopoly market setting, going against the popular economic philosophy/intuition that increased amounts of end-user data signals in a market improves utilitarian social welfare. The primary reason behind our result is the significant negative externality (via user signal correlations) generated by privacy breaches in the information market that cannot be cancelled out via market equilibrium prices handed over to the users for their information. 

\section{Acknowledgement}
This work NSF-supported under grants CNS-1616575, CNS-1939006, CNS-2012001, and ARO W911NF1810208.
\bibliographystyle{unsrt}
\bibliography{scibib,scibib1,references}

\begin{thebibliography}{10}

\bibitem{posner1}
Richard Poser.
\newblock The right of privacy.
\newblock {\em Georgia Law Review}, 12(3), 1978.

\bibitem{posner2}
Richard Poser.
\newblock The economics of privacy.
\newblock {\em American Economic Review}, 71(2).

\bibitem{stigler}
George Stigler.
\newblock An introduction to privacy in economics and politics.
\newblock {\em Journal of Legal Studies}, 9(4), 1978.

\bibitem{Laudon:1996:MP:234215.234476}
Kenneth~C. Laudon.
\newblock Markets and privacy.
\newblock {\em Commun. ACM}, 39(9):92--104, September 1996.

\bibitem{acquisti2016economics}
Alessandro Acquisti, Curtis Taylor, and Liad Wagman.
\newblock The economics of privacy.
\newblock {\em Journal of Economic Literature}, 54(2):442--92, 2016.

\bibitem{odlyzko}
Andrew Odlyzko.
\newblock Privacy, economics, and price discrimination on the internet.
\newblock {\em Economics of Internet Security (Eds. Jean Camp, Stephen Lewis},
  2003.

\bibitem{samuelson2000privacy}
Pamela Samuelson.
\newblock Privacy as intellectual property?
\newblock {\em Stanford law review}, pages 1125--1173, 2000.

\bibitem{schwartz2003property}
Paul~M Schwartz.
\newblock Property, privacy, and personal data.
\newblock {\em Harv. L. Rev.}, 117:2056, 2003.

\bibitem{posner2018radical}
Eric~A Posner and E~Glen Weyl.
\newblock {\em Radical markets: Uprooting capitalism and democracy for a just
  society}.
\newblock Princeton University Press, 2018.

\bibitem{varian2009economic}
Hal~R Varian.
\newblock Economic aspects of personal privacy.
\newblock In {\em Internet policy and economics}, pages 101--109. Springer,
  2009.

\bibitem{farboodi2019big}
Maryam Farboodi, Roxana Mihet, Thomas Philippon, and Laura Veldkamp.
\newblock Big data and firm dynamics.
\newblock In {\em AEA papers and proceedings}, volume 109, pages 38--42, 2019.

\bibitem{pal2019privacy}
Ranjan Pal and Jon Crowcroft.
\newblock Privacy trading in the surveillance capitalism age viewpoints
  on'privacy-preserving'societal value creation.
\newblock {\em ACM SIGCOMM Computer Communication Review}, 49(3):26--31, 2019.

\bibitem{calo2015privacy}
Ryan Calo.
\newblock Privacy and markets: a love story.
\newblock {\em Notre Dame L. Rev.}, 91:649, 2015.

\bibitem{pal2020preference}
Ranjan Pal, Jon Crowcroft, Yixuan Wang, Yong Li, Swades De, Sasu Tarkoma,
  Mingyan Liu, Bodhibrata Nag, Abhishek Kumar, and Pan Hui.
\newblock Preference-based privacy markets.
\newblock {\em IEEE Access}, 8:146006--146026, 2020.

\bibitem{acemoglu2019too}
Daron Acemoglu, Ali Makhdoumi, Azarakhsh Malekian, and Asuman Ozdaglar.
\newblock Too much data: Prices and inefficiencies in data markets.
\newblock Technical report, National Bureau of Economic Research, 2019.

\bibitem{laoutaris2019online}
Nikolaos Laoutaris.
\newblock Why online services should pay you for your data? the arguments for a
  human-centric data economy.
\newblock {\em IEEE Internet Computing}, 23(5):29--35, 2019.

\bibitem{sarwate2013signal}
Anand~D Sarwate and Kamalika Chaudhuri.
\newblock Signal processing and machine learning with differential privacy:
  Algorithms and challenges for continuous data.
\newblock {\em IEEE signal processing magazine}, 30(5):86--94, 2013.

\bibitem{topkis1978minimizing}
Donald~M Topkis.
\newblock Minimizing a submodular function on a lattice.
\newblock {\em Operations research}, 26(2):305--321, 1978.

\bibitem{wang2016value}
Weina Wang, Lei Ying, and Junshan Zhang.
\newblock The value of privacy: Strategic data subjects, incentive mechanisms
  and fundamental limits.
\newblock In {\em ACM SIGMETRICS Performance Evaluation Review}, volume~44,
  pages 249--260. ACM, 2016.

\bibitem{burkschat2014reversed}
Marco Burkschat and Nuria Torrado.
\newblock On the reversed hazard rate of sequential order statistics.
\newblock {\em Statistics \& Probability Letters}, 85:106--113, 2014.

\end{thebibliography}

\end{document}